\documentclass[traditabstract]{aa}

\usepackage{graphicx,amssymb,amsmath,multirow,gensymb,lscape}
\usepackage{enumitem}
\usepackage{natbib}
\newcommand{\Jybeam}{\mbox{Jy beam$^{-1}$}}
\newcommand{\vol}{\mbox{cm$^{-3}$}} 
\newcommand{\cden}{\mbox{cm$^{-2}$}} 
\newcommand{\um}{\mbox{$\mu$m}}
\newcommand{\ammonia}{\mbox{NH$_{3}$}}
\newcommand{\Msun}{\mbox{M$_{\odot}$}}
\newcommand{\cmg}{\mbox{cm$^2$ g$^{-1}$}}

\newcommand{\NHH}{\mbox{N(H$_2$)}}

\begin{document}

\title{Dust emissivity in the star-forming filament OMC 2/3\thanks{Table 1 and a FITS version of the GISMO 2 mm observations are available in electronic form at the CDS via anonymous ftp to cdsarc.u-strasbg.fr (130.79.128.5) or via http://cdsweb.u-strasbg.fr/cgi-bin/qcat?J/A+A/}}

\author{S. I. Sadavoy\inst{1} \and
	 A. M. Stutz\inst{1} \and
	 S. Schnee\inst{2} \and
	 B. S. Mason\inst{2} \and
	 J. Di Francesco\inst{3,4} \and
	 R. K. Friesen\inst{5} 
	 }
\institute{Max-Planck-Institut f\"{u}r Astronomie (MPIA), K\"{o}nigstuhl 17, D-69117 Heidelberg, Germany \and
	National Radio Astronomy Observatory, 520 Edgemont Road, Charlottesville, VA 22903, USA \and
	National Research Council Canada, 5071 West Saanich Road Victoria, BC V9E 2E7, Canada \and
	Department of Physics \& Astronomy, University of Victoria, PO Box 3055 STN CSC, Victoria, BC V8W 3P6, Canada \and
	Dunlap Institute for Astronomy and Astrophysics, University of Toronto, 50 St. George Street, Toronto M5S 3H4, Ontario, Canada 
		}


\date{Received 14 September 2015; accepted 22 January 2016}

\abstract {We present new measurements of the dust emissivity index, $\beta$, for the high-mass, star-forming OMC 2/3 filament.  We combined $160-500$ \um\ data from \emph{Herschel} with long-wavelength observations at 2 mm and fit the spectral energy distributions across a $\simeq$ 2 pc long, continuous section of OMC 2/3 at 15000 AU (0.08 pc) resolution.  With these data, we measured $\beta$ and reconstructed simultaneously the filtered-out large-scale emission at 2 mm.  We implemented both variable and fixed values of $\beta$, finding that $\beta = 1.7 - 1.8$ provides the best fit across most of OMC 2/3.  These $\beta$ values are consistent with a similar analysis carried out with filtered \emph{Herschel} data.  Thus, we show that $\beta$ values derived from spatial filtered emission maps agree well with those values from unfiltered data at the same resolution.  Our results contradict the very low $\beta$ values ($\sim$ 0.9) previously measured in OMC 2/3 between 1.2 mm and 3.3 mm data, which we attribute to elevated fluxes in the 3.3 mm observations.  Therefore, we find no evidence of rapid, extensive dust grain growth in OMC 2/3. Future studies with \emph{Herschel} data and complementary ground-based long-wavelength data can apply our technique to obtain reliable determinations of $\beta$ in nearby cold molecular clouds.
}


\titlerunning{Dust emissivity in the star-forming filament OMC 2/3}
\authorrunning{Sadavoy et al.}
\maketitle

\section{Introduction\label{Intro}}

Dust grains are excellent tracers of mass and structure in molecular clouds.  Thermal emission from dust can characterize structures over various scales, from the diffuse cloud to the dense, star-forming cores \citep[e.g.,][]{difran08, Enoch09, Andre10, StutzKainulainen15}.  Nevertheless, the conversion of thermal dust emission to mass is nontrivial, mainly because of uncertain dust emissivities.  Most studies assume a single power-law distribution for the dust emissivity, $\nu^{\beta}$, resulting in masses that are uncertain by factors of a few \citep[e.g.,][]{Henning95, Shirley11}.  

The dust emissivity index, $\beta$, represents the efficiency at which dust grains radiate at long wavelengths, where the value $\beta = 2$ is expected for bare dust grains in the interstellar medium \citep[e.g.,][]{DraineLee84}.  This efficiency, however, will evolve with density and temperature. For example, dust grains in cold, dense cores are likely to coagulate, \citep[e.g.,][]{Ossenkopf94, Ormel11}, leading to values of $\beta < 2$ toward molecular clouds and dense cores \citep[e.g.,][]{Shirley11,Planck_clouds_beta, Sadavoy13} and values of $\beta < 1$ toward protoplanetary disks  \citep[e.g.,][]{BeckwithSargent91, Wright15}.

The Orion molecular cloud (OMC) 2/3 region is an active, star-forming filament with a rich population of young stars \citep{PetersonMegeath08}.   OMC 2/3 also represents one of the closest sites of high-mass star formation \citep[$\sim 420$ pc;][]{Menten07}, making this region an interesting target to compare with nearby low-mass molecular clouds.   Indeed, many previous studies have examined its dust properties including the dust emissivity index, $\beta$.   \citet{Chini97} used data between $350-2000$ \um\ to determine $\beta$ toward nine cores in OMC 2/3.  For a reasonable range of assumed temperatures, \citet{Chini97} found that most of their sources were well-fit with $\beta = 2$, although several had best-fit values of $\beta \lesssim 1.5$.  Similarly, \citet{JohnstoneBally99} used SCUBA 450 \um\ and 850 \um\ ratios to characterize $\beta$ across the entire Orion integral shaped filament.  For $T > 10$ K, their data agreed well with $1 < \beta < 2$, although their results could not rule out lower $\beta$ indices at higher temperatures.

While these earlier studies found $\beta$ values that agreed well with theoretical predictions for dust grains within molecular clouds and cores,  a recent study by \citet{Schnee14} found much lower $\beta$ values.  \citet{Schnee14} used ratios of 1.2 mm and 3.3 mm observations with \ammonia-derived temperatures \citep[from][]{Li13} to determine $\beta$ in OMC 2/3 for scales up to $\sim 0.1$ pc.  They found $\beta= 0-2$ throughout the OMC 2/3 filament, with a median value of $\beta \simeq 0.9$ and values of $\beta \lesssim 1$ toward   most of the dense cores.  These $\beta$ indices are considerably lower than earlier measurements and are  usually associated with millimeter-size dust grains that are primarily found in disks \citep[e.g.,][]{BeckwithSargent91, WilliamsCieza11} and not with the micron-size dust grains associated with filaments and cores \citep[e.g.,][]{Pagani10, Testi14}.   These results suggest that extraordinary processes may be occurring in OMC 2/3 that produce significant, wide-spread grain growth on $\sim 0.1$ pc scales.  Evidence for such rapid grain evolution has not been before detected.  

To explore the possibility of rapid dust grain growth in OMC 2/3, we combined $160-500$ \um\ observations from the \emph{Herschel} Space Observatory with 2 mm observations from the IRAM 30 m telescope to determine $\beta$ on larger scales than in \citet{Schnee14}.   The 2 mm data are necessary to constrain fits to spectral energy distributions (SEDs) because of the degeneracy between temperature and $\beta$ \citep{Shetty09, Ysard12, Juvela13}.  With the \emph{Herschel} and 2 mm observations, we obtained an independent measure of $\beta$ in OMC 2/3.

\section{Data}\label{data}

\subsection{Herschel}\label{herschelObs}

The Orion molecular cloud was observed as part of the \emph{Herschel} Gould Belt Survey \citep{Andre10}.  The cloud was mapped in the PACS/SPIRE parallel mode to observe the cloud simultaneously at 70 \um, 160 \um, 250 \um, 350 \um, and 500 \um.  We used the $160-500$ \um\ maps presented in \citet{StutzKainulainen15} and \citet{StutzGould15}, which were made by processing the Level 1 map products of the \emph{Herschel} Science Archive with \emph{scanamorphos} \citep[version 24.0,][]{Roussel12}.  The \emph{Herschel} data were corrected for zero-point fluxes using IRAS and \emph{Planck} data \citep[e.g., following][]{Bernard10}. 

\subsection{IRAM 30m telescope}\label{gismoObs}

The OMC 2/3 complex was observed at 2 mm with GISMO \citep{GISMO} at the IRAM 30m telescope on 5 April 2014 in good weather ($\tau_{225 GHz} \lesssim 0.3$).  We used orthogonal $\sim 7$\arcmin\ scans at angles of $\pm\ 30$\degree\ to produce a final map of $\sim 7\arcmin \times 22\arcmin$ at a 40 arcsec s$^{-1}$ scan rate.   The total on-sky time was $\sim 1$ hour.  A larger map ($\sim 8\arcmin \times 34\arcmin$) that includes the OMC 1 region was observed in the same manner on 10 April 2014 in poor weather conditions ($\tau_{225 GHz} \sim 0.5$) for an on-sky time of $\sim 1$ hour.  Additional observations taken at low elevation ($< 30$\degree) and high airmass on 9 April 2014 were not used in this analysis.    

The scans of 5 and 10 April were reduced together using CRUSH \citep{Kovacs08} vers. 2.22 with the ``faint'' and ``extended'' modes and extra iterations.   With these parameters, we expect to recover emission on $\sim 2$\arcmin\ scales to 90\%\ (A. Kov{\'a}cs 2015, private communication; see also Appendix \ref{appendixFT}).  For the OMC 2/3 data, we obtain an 1 $\sigma$ rms sensitivity of $\sim 2.5$ m\Jybeam\ for a $\sim 21$\arcsec\ FWHM beam.  For the OMC 1 complex, which was only observed on 10 April, the 1 $\sigma$ rms sensitivity is $\sim 4$ m\Jybeam.  We use only the OMC 2/3 data at 2 mm hereafter.

\section{Results}\label{results}

\begin{figure}[h!]
\includegraphics[scale=0.46]{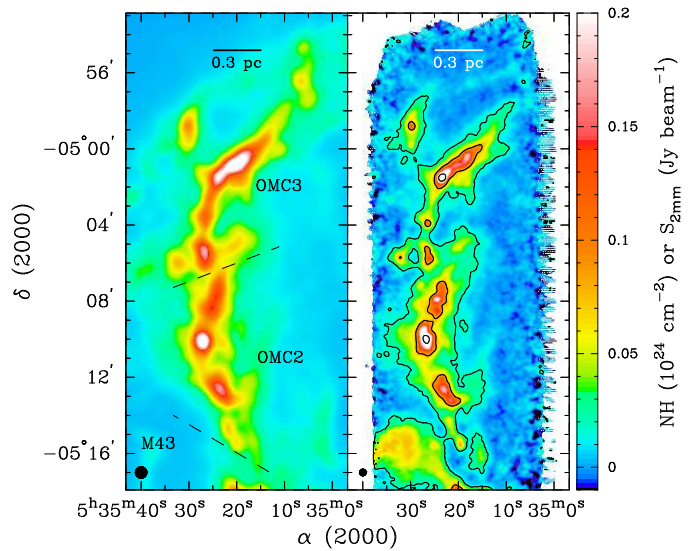}
\caption{Observations of OMC 2/3.  \emph{Left:}  Column densities from \citet{StutzKainulainen15} at $\sim 36$\arcsec\ resolution and assuming a fixed dust opacity law.  The complexes are separated by dashed lines based on \citet{Chini97}.  The \ion{H}{ii} region, M 43, is also labeled.  \emph{Right:} IRAM 2 mm observations at 21\arcsec\ resolution.  Contours show 2 mm flux density levels of 0.02, 0.08, 0.3, and 1 \Jybeam.  Respective beams are shown in the the lower left corners. }\label{gismo_obs}
\end{figure}

Figure \ref{gismo_obs} compares the \emph{Herschel}-derived column density map of OMC 2/3 from \citet{StutzKainulainen15} with the 2 mm observations from GISMO.  The column densities were determined from SED fits to \emph{Herschel} data alone, assuming a fixed dust opacity law based on the models of \citet{Ossenkopf94}. In general, the two maps agree well, suggesting that the \emph{Herschel}-derived column densities trace the locations of dense material well.  The \ion{H}{ii} region M 43, however, is a notable exception where bright ($> 3\ \sigma$), extended emission at 2 mm does not correspond to an increase in column density from the \emph{Herschel} data.  The 2 mm continuum is likely enhanced by free-free emission from the \ion{H}{ii} region.  We exclude M 43 from further analysis in this study.

For the the OMC 2/3 filament, we convolved the \emph{Herschel} and 2 mm data to a common resolution of 36\arcsec.3 and a common grid of 14\arcsec\ pixels, corresponding to the 500 \um\ data.   Dust temperatures and $\beta$ indices were determined from fitting the observed SEDs with the modified blackbody function,
\begin{equation}\label{eqBB}
I_{\nu} = \kappa_0(\nu/\nu_0)^{\beta}B_{\nu}(T)\Sigma,
\end{equation}
where $\kappa_0$ is the dust opacity at the reference frequency $\nu_0$, $B_{\nu}$ is the blackbody function at temperature $T$, and $\Sigma$ is the gas mass column density.  For this study, we assumed $\kappa_0 = 0.1$ \cmg\ and $\nu_0 = 1$ THz in agreement with the \emph{Herschel} Gould Belt Survey \citep{Andre10}.

The SED fitting is nontrivial.  Since the 2 mm observations were obtained at a ground-based facility, these data are filtered on scales larger than a few arcmin, whereas the \emph{Herschel} data recover emission on these scales.  Thus, the pixel-by-pixel SEDs will be missing emission at 2 mm relative to the \emph{Herschel} data.  To fit the \emph{Herschel}+2mm SEDs, we must either recover the large-scale diffuse emission at 2 mm or similarly filter out the large-scale emission from the \emph{Herschel} data.  Here, we examine both possibilities.

\subsection{Unfiltered data}\label{largeScaleSection}

In our first approach, we recovered the large-scale structure at 2 mm using a grid of offsets to represent the filtered-out large-scale emission.  For each assumed offset value, we fitted SEDs over subregions of $9 \times 9$ pixels ($2.1\arcmin \times 2.1\arcmin$) where we expect the extended emission to be smooth (see Appendix \ref{appendixFT}).  We then identified the best-fit offsets for various assumed values of $\beta$.  Offsets that deviate significantly from the true value of the extended emission will be poorly fit with a modified blackbody function to the $160-2000$ \um\ SED, whereas those offset values that match the true value of the extended emission well will produce better fits.  By minimizing the global $\chi^2$ across all pixels in a $9\times9$ pixel subregion, we estimated the filtered-out large-scale emission toward each pixel.  For more details, see Appendix D in \citet{Sadavoy13}.

We applied our technique with $\beta$ as a free parameter and with fixed values of $\beta = 1.3-2.1$.  For the $\beta$ trials, we adopted a grid of offsets from $0-0.2$ \Jybeam\ for $\beta > 1.4$ and a coarser grid of $0-0.4$ \Jybeam\ for $\beta \le 1.4$.   Moreover, we fitted only those pixels with $S_{2mm} > 5\ \sigma$ to ensure that the observed 2 mm emission can be used as a constraint in the SED fit.

\begin{figure}[h!]
\includegraphics[scale=0.47]{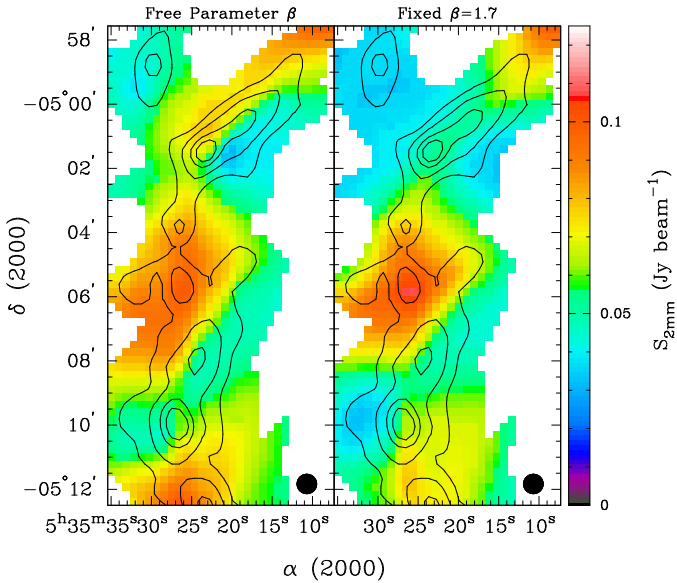}
\caption{Recovered large-scale emission (``offsets'') for $\beta$ is a free parameter (left) and when $\beta = 1.7$ (right).  Contours show flux densities of 0.08, 0.16, 0.32, and 0.48 \Jybeam\ from the \emph{observed} 2 mm map at 36\arcsec\ resolution (black circle).  Note that the structures in the offset map have scales $> 5$ arcmin.}\label{offset_comp}
\end{figure}

Each $\beta$ trial resulted in different best-fit offsets, $S_{off}$.  For example, $\beta = 2.1$ produced only best-fit offsets of $S_{off} \lesssim 0.03$ \Jybeam\ throughout OMC 2/3, whereas $\beta =1.3$ gave only best-fit offsets of $S_{off} > 0.06$ \Jybeam.  Nevertheless, we see common features in most cases. Figure \ref{offset_comp} shows the best-fit offset results for the cases of  $\beta$ as a free parameter and $\beta = 1.7$.  In both maps, we find that $S_{off} \sim 0.3 - 0.1$ \Jybeam\ across OMC 2/3 with typical uncertainties of $\lesssim 20$\%.   Both maps also give similar, large-scale clumpy structures.  These features are seen in many of the $\beta$ trials and most likely reflect the missing large-scale structure toward OMC 2/3 at 2 mm.

\begin{figure}[h!]
\includegraphics[scale=0.55]{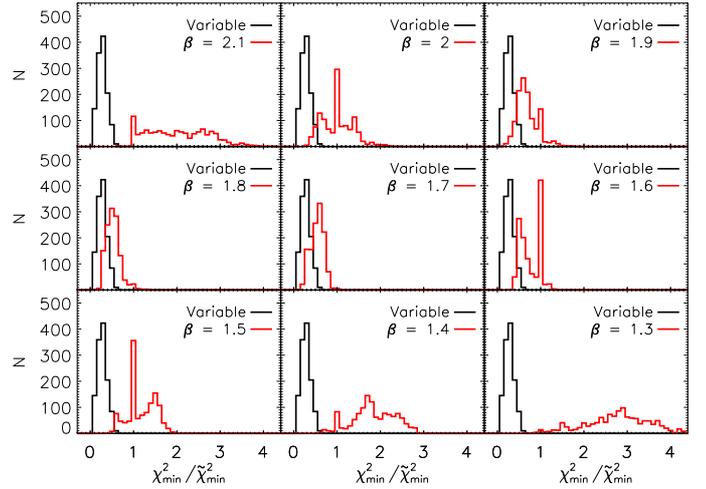}
\caption{Minimum $\chi^2$ values ($\chi^2_{min}$) from fits to each pixel scaled by the median lowest $\chi^2$ value ($\widetilde{\chi}^2_{min}$) from all ten trials for that pixel.  The plots compare fixed values of $\beta$ (red histograms) with the case of $\beta$ as a free parameter (black histograms). }\label{beta_comp}
\end{figure}

We identified the best-fit offsets for each $\beta$ trial, but not all $\beta$ trials produced reliable measurements.   Figure \ref{beta_comp} compares the best-fit results from several $\beta$ trials.  The histograms show the minimum total $\chi^2$ value ($\chi^2_{min}$) for each pixel in each $\beta$ trial normalized by the median lowest $\chi^2$ value ($\widetilde{\chi}^2_{min}$) for that pixel across all presented trials.  With this scaling, values $< 1$ performed better than the median and values $> 1$ performed worse.  We find that $\beta \gtrsim 2$ and $\beta \lesssim 1.5$ do not fit our observations well, whereas $\beta=1.8$ and $\beta=1.7$ generally performed much better than the median value.  The 2 mm band is very sensitive to slight changes in $\beta$.  For example, a change from $\beta = 2$ to $\beta = 1.8$ at 20 K increases the 2 mm emission by 46\%, whereas the \emph{Herschel} bands differ by $\lesssim$ 10\%.

\begin{figure}[h!]
\includegraphics[scale=0.55]{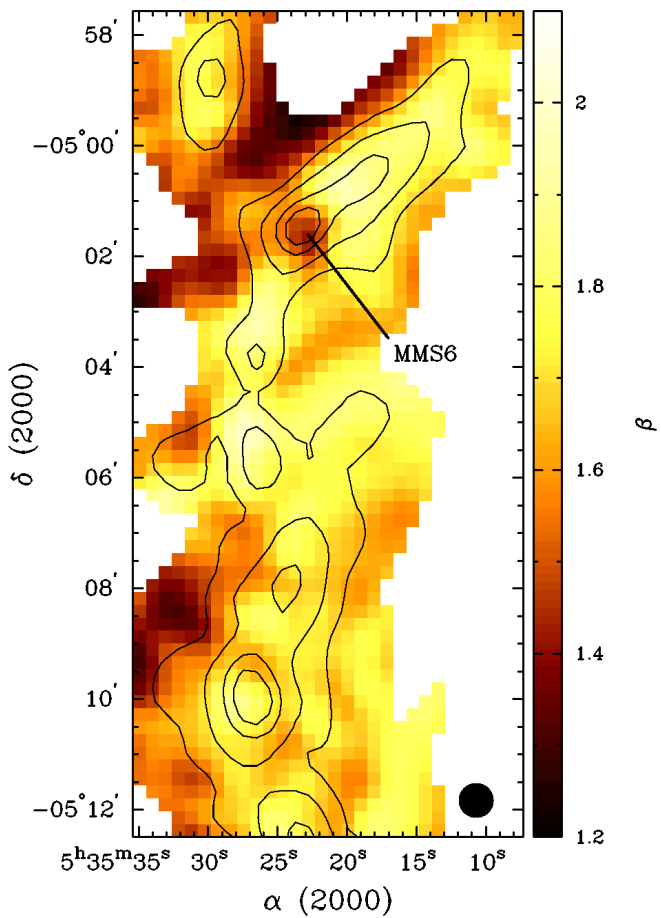}
\caption{Dust emissivity map of OMC 2/3 when $\beta$ is a free parameter.  Contours show observed 2 mm flux densities of 0.08, 0.16, 0.32, and 0.48 \Jybeam\ at 36\arcsec\ resolution (black circle).  The source MMS 6 \citep{Chini97} is also highlighted.}\label{beta_free_map}
\end{figure}

Figure \ref{beta_free_map} shows the distribution of $\beta$ across OMC 2/3 from the trial with $\beta$ as a free parameter.  We find a median value of $\beta = 1.7$, in agreement with the fixed $\beta$ results (Fig. \ref{beta_comp}).  Nevertheless, Fig. \ref{beta_free_map} shows $\beta \simeq 1.3$ toward the map edges and $\beta \simeq 1.4$ for MMS 6.   Along the map edges, the dust temperatures were $> 25$ K (see Sect. \ref{comparisonSection}), and at these temperatures, SEDs are not constrained well with a shortest wavelength of 160 \um.  The MMS 6 source has $T \simeq 20$ K, however, so its SED fits are still expected to be reliable.   Assuming offset uncertainties of 20\%\ and calibration uncertainties of 10\%\ in each band (with the SPIRE calibration errors correlated), we conducted a Monte Carlo analysis of the measurement uncertainties in $\beta$.  We found $\sigma_{\beta} \simeq 0.1$, indicating that the lower $\beta$ values toward MMS 6 are significant within the measurement uncertainties.

\subsection{Filtered data}\label{filterSection}

In our second approach, we filtered the \emph{Herschel} data to match the spatial scales recovered  in the GISMO 2 mm data.  We adopted a similar Fourier-based technique as \citet{Wang15} to remove the large-scale structures from our \emph{Herschel} data.  Similar to a highpass filter, we created a frequency domain filtering mask by fitting an exponential function to the 2 mm amplitude profile.  This mask was designed to retain the small-scale structures and suppress the large-scale emission in a similar manner as the GISMO data.  We applied this mask to the \emph{Herschel} data in the frequency domain to produce our filtered \emph{Herschel} maps.  See Appendix \ref{appendixFT} for more details.

We fitted the filtered \emph{Herschel}+GISMO SEDs directly using Eq. \ref{eqBB}.  We processed only those pixels with $> 5\ \sigma$\ in all wavebands to avoid artifacts in the filtering (see Appendix \ref{appendixFT}).   Figure \ref{beta_filt} shows the distribution of $\beta$ in OMC 2/3 from the filtered data.  The filtered data produced a wider range in $\beta$ values than the unfiltered data, with $\beta = 0.7 - 2.1$ across OMC 2/3.  The locations with $\beta \lesssim 1$, however, are exclusively toward the edges of the filament and most likely result from the filtering process (see Appendix \ref{appendixFT}).  Excluding these data points, the filtered data give a median value of $\beta \simeq 1.6$.  Figure \ref{beta_filt} also shows the ratio of the $\beta$ values obtained from the filtered and unfiltered data.  Except for the filament edges, where we have possible filtering artifacts, the two datasets agree well.  In general, we find that the filtered and unfiltered $\beta$ indices agree within $\sim 20$\%, similar to the measurement uncertainties, with lower indices (by a factor of $\gtrsim 0.7$) obtained from the filtered data.

\begin{figure}[h!]
\centering
\includegraphics[trim=0pt -0.9cm 0cm 1mm,clip=true,scale=0.585]{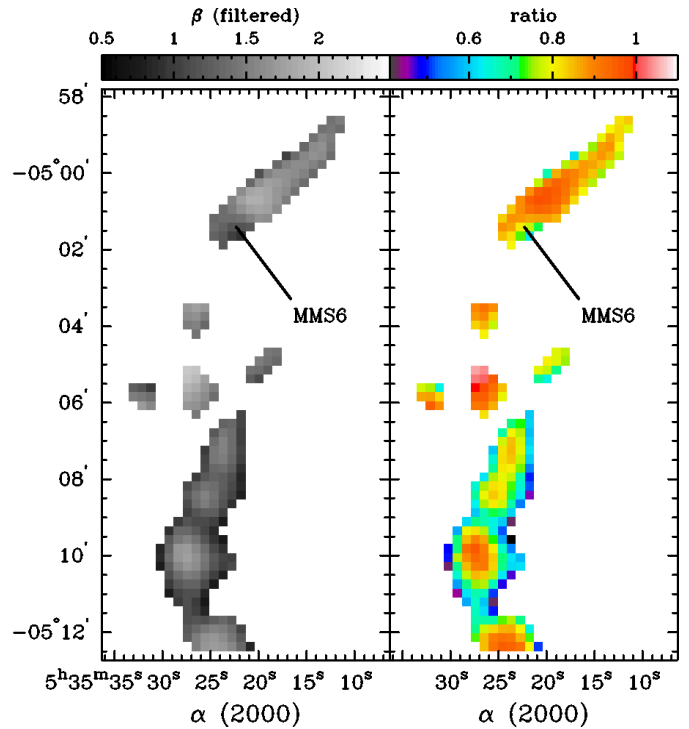}\\[-5mm]
\caption{Best-fit $\beta$ values from Fourier-filtered \emph{Herschel} data and our observed GISMO 2 mm observations (left) and the ratio of these $\beta$ values to the $\beta$ values from the unfiltered data in Sect. \ref{largeScaleSection}.}\label{beta_filt}  
\end{figure}  

Similar to the unfiltered case (Sect. \ref{largeScaleSection}) there are also hints of lower $\beta$ indices toward the MMS 6 source.  Figure \ref{beta_filt} covers MMS 6 only partially because this source was coincident with filtering artifacts, but there is sufficient structure away from these artifacts to suggest that $\beta \simeq 1.2$ toward MMS 6.  Thus, both methods suggest unique dust emissivities toward this one source.

\section{Discussion}\label{discussion}

We found values of $\beta = 1.7-1.8$ with our unfiltered analysis and $\beta \simeq 1.6$ with our filtered analysis across the OMC 2/3 filament.  The two methods agree within their uncertainties, suggesting that spatially filtering our observations did not greatly affect the derived values of $\beta$.   We also found that the $\beta$ indices toward the MMS 6 core were consistently lower than those found along the main filament.  MMS 6 has $\beta \simeq 1.4$ and $\beta \simeq 1.2$ with the unfiltered and filtered data, respectively.  These unique values suggest that MMS 6 may have unique dust properties (see also, Sect. \ref{grainSection}).

Our $\beta$ distributions agree well with the results from \citet{Chini97} and \citet{JohnstoneBally99}.  Those studies similarly found $1 < \beta < 2$ for assumed temperatures typical of star-forming regions.  In particular,  \citet{Chini97} also found evidence of lower $\beta$ indices ($\beta \lesssim 1.5$) toward the MMS 6 source, in agreement with our results \citep[see also,][]{Lis98}.  In comparison, \citet{Schnee14} found much lower $\beta$ values compared to the results in this study and in the earlier studies of \citet{Chini97} and \citet{JohnstoneBally99}.  For example, they found a median value of $\beta \simeq 0.9$ and values of $\beta \simeq 0.5$ toward MMS 6.  Thus, the very low $\beta$ values from \citet{Schnee14} appear to be an outlier.  We investigate the reason for this $\beta$ discrepancy in the following sections.

\subsection{Dust emissivity in dense cores}\label{sourceSection}

As an additional test of $\beta$ in OMC 2/3, we examined the compact, dense cores themselves.   We extracted source fluxes from the \emph{Herschel}+GISMO data using the extraction code \emph{getsources} \citep{getsources, getsources+filaments}.  In brief, \emph{getsources} identifies structures over various scales through spatial decompositions at each wavelength and then combines the data to measure source properties based on information from all wavelengths.  In this way, higher resolution information is used to identify structures at lower resolution.  Moreover, \emph{getsources} is designed to identify and extract intensity peaks over a varied background, and as such, the algorithm can be used on other bolometer instruments.  Nevertheless, since the GISMO data are filtered, we restricted the spatial decompositions to $< 2.5$\arcmin.  We selected 17 reliable detections based on conservative selection criteria: (1) signal-to-noise ratios (S/N) of S/N $> 3$ in at least three of the \emph{Herschel}+GISMO bands and (2) no flags indicative of unreliable detections, such as low S/N or few detections in the multi-scale decompositions.  A complete catalog of sources in Orion will be made available from the \emph{Herschel} Gould Belt Survey.

Table \ref{sourceTable} lists the 17 reliable detections.  Nearly all sources were also observed at 1.3 mm by \citet{Chini97} and are labeled following their naming scheme.  For those sources that were not observed by \citet{Chini97}, we used the source numbers from the 850 \um\ SCUBA survey of \citet{Nutter07}.  Table \ref{sourceTable} also lists the properties of our sources.  We measured the source properties from fitting their SEDs using Eq. \ref{eqBB} with a mean molecular weight of $\mu = 2.8$ in two ways: (1) fitting for the dust temperature, and (2) adopting a fixed dust temperature given by the \ammonia\ kinetic gas temperature from \citet{Li13}.  For dense objects ($\gtrsim 10^6$ \vol), the gas and dust are expected to be coupled, resulting in similar temperatures \citep[e.g.,][]{Young04, Ceccarelli07}.  Several sources in Table \ref{sourceTable}, however, have gas and dust temperatures that differ by $> 5$ K and may be more tenuous.  Alternatively, the gas and dust may not be tracing the same material (e.g., for protostellar objects, \ammonia\ may trace the cool outer envelope, whereas the dust will trace primarily the warm inner envelope).  The parameter errors correspond to 1 $\sigma$ errors following a Monte Carlo analysis of the SED-fitting within the observational uncertainties.

\begin{table*}
\begin{center}
\caption{Source properties}\label{sourceTable}
\begin{tabular}{lccccccccc}
\hline\hline
					&					&		&		& \multicolumn{3}{c}{Dust Temperatures}	&	\multicolumn{3}{c}{Gas Temperatures\tablefootmark{c}} \\
Name\tablefootmark{a}	& HOPS\tablefootmark{b}	&	RA	&	Dec	& $T_{dust}$	&	$\beta$	&	$M$		&   $T_{gas}$	& $\beta$		& 	$M$	   \\
					&					& (J2000)	& (J2000)	&	(K)		&			& 	(\Msun)	&	(K)		&			& 	(\Msun) \\
\hline
NW167 	& 096	& 5:35:30.0 &  -4:58:48 &    17.7 $\pm$  2.2 	&    1.8 $\pm$ 0.2 	&   0.8 $\pm$  0.3 & $\cdots$ &  $\cdots$ 	&  $\cdots$ \\  
NW165 	& 383	& 5:35:30.0 &  -4:59:46 &    20.5 $\pm$  2.5  	&    1.8 $\pm$ 0.1 	&   0.2 $\pm$  0.1 & $\cdots$ &  $\cdots$ 	&  $\cdots$ \\  
MMS 2 	& 092	& 5:35:18.6 &  -5:00:31 &    14.6 $\pm$  1.1 	&    2.0 $\pm$ 0.1 	&   4.4 $\pm$  1.4 	&       16 &    1.9 $\pm$ 0.1 &    3.2 $\pm$ 0.2 \\  
MMS 4 	& 089$^{d}$ & 5:35:20.4 &  -5:00:51 &     9.7 $\pm$  0.6 &    2.9 $\pm$ 0.2 	&  14 $\pm$  0.1 	&       14 &    2.0 $\pm$ 0.1 &    2.4 $\pm$ 0.1 \\  
MMS 5 	& 088	& 5:35:22.6 	&  -5:01:14 &    21.5 $\pm$  2.5 &    1.5 $\pm$ 0.1 &   0.6 $\pm$  0.2 &       19 &    1.7 $\pm$ 0.1 &    1.0 $\pm$ 0.1 \\  
MMS 6 	 & 086	& 5:35:23.6 	&  -5:01:31 &    24.9 $\pm$  3.3 &    1.3 $\pm$ 0.1 &   0.8 $\pm$  0.3 &       27 &    1.2 $\pm$ 0.1 &    0.7 $\pm$ 0.1 \\  
MMS 7 	& 084 	& 5:35:26.6 	&  -5:03:57 &    28.0 $\pm$  4.4 &    1.7 $\pm$ 0.1 &   0.3 $\pm$  0.1 &       19 &    2.2 $\pm$ 0.1 &    0.9 $\pm$ 0.1 \\  
MMS 8 	& $\cdots$ & 5:35:26.8 	&  -5:05:18 &    12.3 $\pm$  0.8 &    2.5 $\pm$ 0.1 &   3.5 $\pm$  1.0 &       13 &    2.4 $\pm$ 0.1 &    2.8 $\pm$ 0.2 \\  
MMS 9 	& 078	& 5:35:26.3 	&  -5:05:45 &    15.2 $\pm$  1.3 &    2.1 $\pm$ 0.1 &   1.8 $\pm$  0.6 &       19 &    1.8 $\pm$ 0.1 &    0.8 $\pm$ 0.1 \\  
MMS 10 	& 077$^{d}$ & 5:35:32.4 &  -5:05:49 &    15.8 $\pm$  1.3 &    2.1 $\pm$ 0.1 &   0.9 $\pm$  0.3 & $\cdots$ &           $\cdots$ &           $\cdots$ \\    
FIR 2  	& 068	& 5:35:24.7 	&  -5:08:32 &    14.2 $\pm$  2.0 &    1.9 $\pm$ 0.2 &   0.8 $\pm$  0.4 &       18 &    1.6 $\pm$ 0.1 &    0.4 $\pm$ 0.1 \\  
FIR 3  	& 370	& 5:35:27.8 	&  -5:09:34 &    34.5 $\pm$  6.7 &    1.7 $\pm$ 0.1 &   0.5 $\pm$  0.3 &       28 &    1.9 $\pm$ 0.1 &    1.0 $\pm$ 0.1 \\  
FIR 4  	& 108	& 5:35:27.0 	&  -5:09:59 &    28.5 $\pm$  4.3 &    1.5 $\pm$ 0.1 &   2.1 $\pm$  0.8 &       23 &    1.6 $\pm$ 0.1 &    3.8 $\pm$ 0.2 \\ 
FIR 6b  	& 060	& 5:35:23.6 	&  -5:12:04 &    21.4 $\pm$  2.8 &    1.9 $\pm$ 0.2 &   0.5 $\pm$  0.2 &       17 &    2.2 $\pm$ 0.1 &    1.3 $\pm$ 0.1 \\  
FIR 6a  	& $\cdots$ & 5:35:23.5 	&  -5:12:38 &    17.0 $\pm$  1.6 &    1.9 $\pm$ 0.1 &   2.8 $\pm$  1.0 &       17 &    1.9 $\pm$ 0.1 &    3.0 $\pm$ 0.2 \\  
FIR 6c 	& 059	& 5:35:21.7 	&  -5:13:14 &    15.6 $\pm$  1.2 &    2.1 $\pm$ 0.1 &   2.2 $\pm$  0.7 &       24 &    1.6 $\pm$ 0.1 &    0.6 $\pm$ 0.1 \\  
NW099 	& 056	& 5:35:19.8 	&  -5:15:35 &    23.3 $\pm$  3.2 &    1.9 $\pm$ 0.1 &   0.5 $\pm$  0.2 &       15 &    2.6 $\pm$ 0.1 &    2.2 $\pm$ 0.1 \\  
\hline\\[-8mm]
\end{tabular}
\end{center}
\tablefoot{
\tablefoottext{a}{Source names with ``MMS'' or ``FIR'' are from \citet{Chini97} and source names with ``NW'' are from \citet{Nutter07}.}
\tablefoottext{b}{Source identifications from the \emph{Herschel} Orion Protostar Survey \citep[HOPS;][]{Furlan16}.}
\tablefoottext{c}{Results from SED-fitting with the dust temperature fixed to the kinetic gas temperature from \citet{Li13}.}
\tablefoottext{d}{Tentative association with HOPS source.} 
}
\end{table*}	

We find a wider range in $\beta$ values ($1.2 \lesssim \beta \lesssim 3$) for the cores compared to the $\beta$ values across the OMC 2/3 filament from the unfiltered and filtered data, where $\beta$ was more uniform.  Since large variations in $\beta$ suggest different dust properties, Table \ref{sourceTable} suggests that the dense cores may have very different dust grain properties.  Moreover, these different grain populations are very localized, as they are not noticeable in the analyses of the larger-scale filament.

Table \ref{sourceTable} shows MMS 6 has the lowest value of $\beta \sim 1.2-1.3$, whereas all remaining sources have $\beta > 1.5$.  These lower $\beta$ values for MMS 6 agree well with our previous results (see Figs. \ref{beta_free_map} and \ref{beta_filt}) and emphasize that this object may be unique.  For the other sources in Table \ref{sourceTable}, we find a wide range of $\beta$ indices and often large differences in $\beta$ from adopting either $T_{dust}$ and $T_{gas}$.  Since most of our sources are protostellar \citep{Megeath12, Furlan16}, our single-temperature modified blackbody functions may be too simplistic to represent the source SEDs accurately.  More extensive SED modeling with a better-sampled SED will improve our understanding of the dust properties.

\subsection{Evidence of dust grain growth?}\label{grainSection}

The dust emissivity index, $\beta$, is often used as an indirect probe of the dust grain population.   Dust grains are most efficient at emitting radiation at wavelengths similar to their size, and as such, larger dust grains should emit more efficiently at long wavelengths than smaller dust grains.  Nevertheless, $\beta$ is sensitive to many additional factors, including the dust composition, structure, and the presence or absence of ice mantles \citep[e.g.,][]{Ossenkopf94, Ormel11}.  Moreover, depending on the physics and model, $\beta$ can remain relatively constant for different dust grain populations \citep{Ormel11}.

Although the link between $\beta$ and dust grain populations is very complex, very low values of $\beta$ (e.g., $\beta \lesssim 1$) are most likely caused by substantial grain growth, whereas $\beta \simeq 2$ most likely reflects less processed dust associated with clouds and filaments \citep[e.g.,][]{Testi14}.  Thus, larger deviations from $\beta = 2$ will still be informative even though we cannot connect small changes in $\beta$ unambiguously with grain evolution.

Using our unfiltered maps, we found $\beta \simeq 1.7$ throughout most of OMC 2/3, with minimum values of $\beta \simeq 1.4$ toward MMS 6 (see Fig. \ref{beta_free_map}).  These values agreed well with the results from the filtered maps, where $\beta \simeq 1.6$ with lower values of $\beta \simeq 1.2$ toward MMS 6.  Thus, the MMS 6 source appears to be unique (see also Table \ref{sourceTable}).  It is also one of the densest objects in OMC 2/3.   Since \citet{Roy13} found a strong link between column density and dust opacity in the southern parts of Orion A (excluding OMC 2/3), dust grain evolution in MMS 6 is expected.  Alternatively, since MMS 6 is a protobinary system \citep{Furlan16},  its lower $\beta$ values might be indicative of a more complicated SED than portrayed by our single-temperature modified blackbody functions \citep[e.g.,][]{Shetty09}. Therefore, while the lower $\beta$ indices in MMS 6 hint at unique dust grain properties, we cannot rule out a broadened SED due to multiple temperature components.

Even if we assume that the lower $\beta$ indices toward MMS 6 are due to dust grain evolution, we find no evidence of $\beta < 1$ either toward the sources or along the filament.  In contrast, \citet{Schnee14} found $\beta < 1$ toward most of the OMC 2/3 filament on scales of 0.1 pc.  Since our observations resolve this 0.1 pc scale ($\sim 50\arcsec$), these low values of $\beta$ should be detectable.  Therefore, we find no evidence of the large dust grains in OMC 2/3 suggested by \citet{Schnee14}.  We discuss the possible reasons for this discrepancy in Sect. \ref{freefreeSection}.

\subsection{Understanding the $\beta$ contradiction}\label{freefreeSection}

The results presented here and in \citet{Schnee14} represent independent measurements of $\beta$ across OMC 2/3 with different datasets and different techniques.   Our two studies are inconsistent.  Moreover, fixed values of $\beta = 0.9$ with our SED-fitting analysis (see Sect. \ref{largeScaleSection}) yielded very poor results, with $\chi^2$ values that well exceeded the $\beta = 1.3$ case (see Fig. \ref{beta_comp}).  Indeed, fixed values of $\beta \lesssim 1.3$ yield subsequently inferior results that disagree with our observations.

\citet{Schnee14} used gas temperatures from \citet{Li13} to determine $\beta$ from their 1.2 mm and 3.3 mm observations, whereas we fitted our SEDs with a line-of-sight average dust temperature.  If these temperatures were considerably different (e.g., due to tracing different material), they might explain the discrepancy in $\beta$.  The gas and dust temperatures, however, showed good general agreement.  For example, the gas temperatures were typically cooler than the dust temperatures by $\lesssim 5$ K, although some of the protostellar sources had larger variations of $\sim 10$ K (see also Table \ref{sourceTable}).  If we were to decrease all our dust temperatures to correspond better to the gas temperatures, $\beta$ would become steeper and deviate more strongly from the results in \citet{Schnee14}.  Therefore, the kinetic gas temperatures cannot account for $\beta \sim 0.9$ in \citet{Schnee14}.

\begin{figure}[h!]
\includegraphics[scale=0.5]{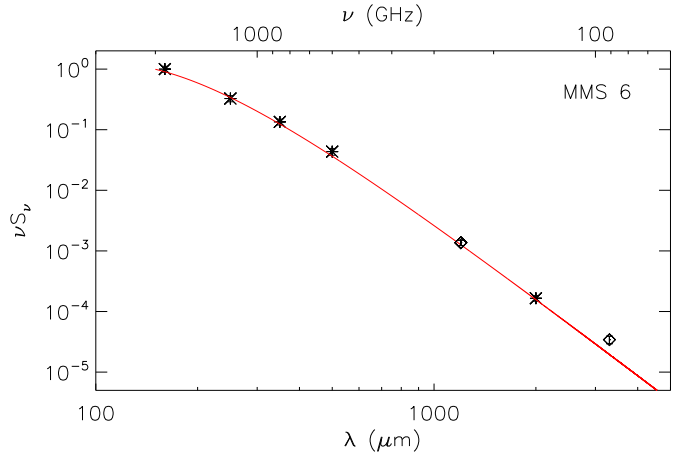}
\caption{Spectral energy distribution for MMS 6.  Stars correspond to \emph{Herschel} and GISMO 2 mm fluxes and diamonds show MAMBO 1.2 mm and MUSTANG 3.3 mm data from \citet{Schnee14} for this source.  The red curve shows the best-fit SED to the \emph{Herschel}+GISMO data alone (see Table \ref{sourceTable}).  Error bars show a 10\%\ flux uncertainty, representative of typical calibration errors. }\label{mms6_sed}
\end{figure}

Alternatively, underestimated ratios of 1.2 mm to 3.3 mm emission would bias the results of \citet{Schnee14} to lower $\beta$ indices.   Figure \ref{mms6_sed} shows the SED for MMS 6 taking our \emph{Herschel}+GISMO data and the 1.2 mm and 3.3 mm fluxes from \citet{Schnee14}.  The best-fit SED (red curve) was determined from fitting the \emph{Herschel}+GISMO data (see Table \ref{sourceTable}) and is thus determined independently of the 1.2 mm and 3.3 mm emission.  The 1.2 mm MAMBO data show excellent agreement with our best-fit SED, whereas the 3.3 mm emission appears to be elevated by roughly a factor of two.  Assuming that the 3.3 mm band was uniformly elevated by a factor of two across all of OMC 2/3, ratios of 1.2 mm to ``corrected'' 3.3 mm emission would yield $\beta \simeq 1.5$.   Such $\beta$ indices agree well with our results using the filtered \emph{Herschel} and GISMO 2 mm data.

The 3.3 mm emission reported by \citet{Schnee14} may have contributions from sources beyond dust, such as molecular line emission, anomalous microwave emission (AME), or free-free emission.  Molecular line contamination in wide continuum bands is restricted mainly to very bright or very broad lines from CO gas \citep[e.g.,][]{Drabek12}.   For MUSTANG, these lines lie outside the bandpass \citep{Schnee14} and will not affect the continuum fluxes.  AME is also expected to be negligible at 3.3 mm.  \citet{Schnee14} used \emph{Planck} observations to estimate the AME toward OMC 2/3 and found it to be well below the thermal dust emission (by over three orders of magnitude) at $\sim 3$ mm.

For free-free emission, \citet{Reipurth99} found 14 sources of radio emission in the OMC 2/3 filament with the VLA in X-band (3.6 cm).  Assuming that the radio emission traces optically thin free-free emission, \citet{Schnee14} determined that the contribution at 3.3 mm would be negligible.   If the free-free emission traced at 3.6 cm is optically thick, however, it might contribute significantly to the 3.3 mm continuum.  Nevertheless, such significant free-free contamination at 3.3 mm would similarly elevate the 2 mm emission by a factor of $\sim 1.2$, which would be noticeable compared to the 1.3 mm data in Fig. \ref{mms6_sed}.  Moreover, the radio sources in \citet{Reipurth99} were generally compact ($\lesssim 8\arcsec$ or $\sim 0.02$ pc), making any contamination at 0.1 pc scales unlikely.    A more detailed analysis of the radio spectrum toward OMC 2/3 is planned to characterize any free-free contributions to the observed millimeter fluxes toward the dense cores and along the filament.

If not from contamination, the elevated 3.3 mm observation could arise if the dust emissivity function deviates from a single power law.  Many laboratory studies have found complicated dust emissivity functions at long wavelengths such that a single $\beta$ value does not describe well the efficiency of these grains \citep[e.g.,][]{Coupeaud11, Paradis11}.  \citet{Meny07} predicted shallower $\beta$ indices at long wavelengths ($> 500$ \um) that are due to changes in the structure of the dust grains themselves.  Thus, the elevated fluxes at 3.3 mm might correspond to a more complicated dust emissivity function.  Alternatively, the 3.3 mm data themselves might have systematic errors from the observations or the data reduction that have not been identified.  Future observations at long wavelengths from larger cameras \citep[e.g., MUSTANG-2;][]{Dicker14} will help address this concern.

\subsection{Comparison to Planck}\label{planckSection}

In Sect. \ref{largeScaleSection}, we recovered the large-scale emission at 2 mm using the \emph{Herschel} data and different assumptions of $\beta$.  Thus, the summation of our observed 2 mm data and the estimated large-scale 2 mm emission should yield the total 2 mm emission of OMC 2/3.  We compared such a total 2 mm map against similar observations from \emph{Planck}\footnote{\emph{Planck} is an ESA science mission with instruments and contributions directly funded by ESA Member States, NASA, and Canada.} observations at 143 GHz from the Full Release vs 2.0.  

We extracted a $3\degree \times 3\degree$ map centered on OMC 1 from the \emph{Planck} Legacy Archive.  Since these data have a spatial resolution of $\sim 7.3$\arcmin\ ($\sim 1$ pc for Orion) at 143 GHz \citep{MeisnerFinkbeiner15}, we convolved our total 2 mm map (for $\beta = 1.7$) to match Plank.  (We note that the \emph{Planck} spatial resolution corresponds to the width of our GISMO observations; see Fig. \ref{gismo_obs}.) We converted both maps to units of MJy sr$^{-1}$ \citep[see][for the \emph{Planck} conversion]{Planck_spectralResponse} and placed all maps onto a common grid.  Compared to the \emph{Planck} 143 GHz data, our total 2 mm map underestimated the 2 mm emission by only $\sim 8$\%\ on average (toward the inner $\sim 7$\arcmin\ of the filament),  whereas the original (uncorrected) 2 mm map underestimated the 2 mm emission by $\sim 43$\% on average.  Thus, our technique recovered the missing large-scale structure from the GISMO 2 mm data well.  

We also compared our $\beta$ indices with those determined by \citet{Planck_foreground}.  We used the $\beta$ maps from the Full Release vs 2.0 available in the \emph{Planck} archive.  These data have a resolution of $7.5\arcmin$, similar to the width of the OMC 2/3 filament.  The \emph{Planck}-derived $\beta$ map shows $1.6 < \beta < 2.0$ and a median of $\beta \simeq 1.7$, with higher $\beta$ values toward the north and south.  Thus, the \emph{Planck}-derived $\beta$ indices of OMC 2/3 agree well with the results from our unfiltered maps.   

With our SED-fitting technique, we can recover the large-scale emission from ground-based data and determine $\beta$ at resolutions superior to those of \emph{Planck} (e.g., $36$\arcsec\ for our GISMO data versus $7.5$\arcmin\ for the \emph{Planck} data at 2 mm).   Alternatively, we can use the approach taken by \citet{Csengeri16}, where the missing large-scale emission in ATLASGAL observations at 870 \um\ was supplanted directly by \emph{Planck} observations.  At 870 \um, the \emph{Planck} data have a resolution of 5$\arcmin$, which corresponds better to the missing large-scale emission in most ground-based facilities (e.g., see Fig. \ref{amp_herschel_b1}).

\subsection{Temperature and column density}\label{comparisonSection}

\begin{figure}[h!]
\centering
\includegraphics[scale=0.62,trim=0pt 0cm 0cm 0mm,clip=true]{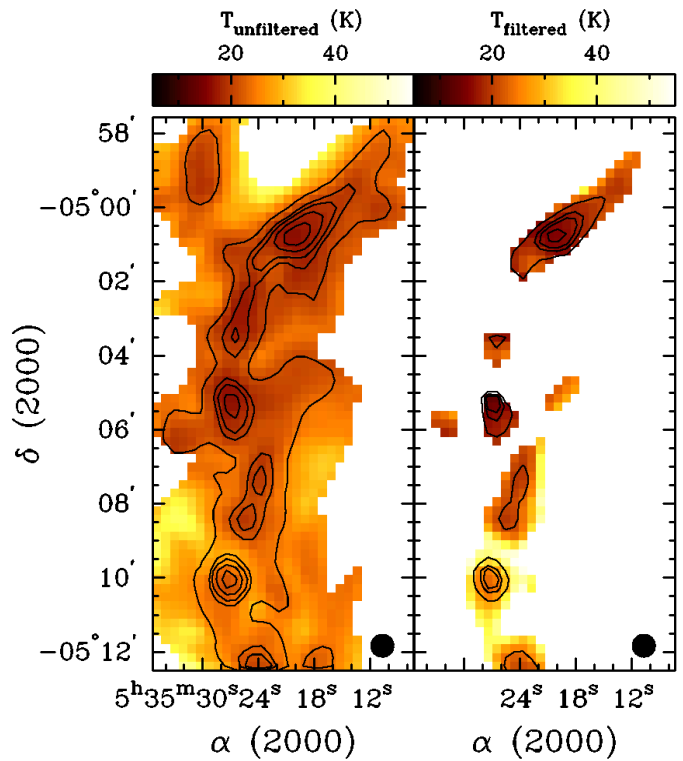}
\caption{Temperature maps of OMC 2/3 using unfiltered data (left) and filtered data (right).  Contours show corresponding H$_2$ column densities at 2, 4.5, 6, 8.5, and 12 $\times 10^{22}$ \cden. For the unfiltered maps, we show the results for $\beta$ as a free parameter. }\label{tempMaps}
\end{figure}

Many studies have used SED fits to examine the temperature and mass structure of molecular clouds with fixed values of $\beta$.  Here, we fitted SEDs with flexible values of $\beta$, which we expect to provide a more reliable measurement of the line-of-sight dust properties.  Figure \ref{tempMaps} shows the temperature maps from our unfiltered and filtered \emph{Herschel}+GISMO analyses with corresponding H$_2$ column density contours.  We find the coolest temperatures ($< 20$ K) generally toward the densest material (e.g., $\NHH > 2 \times 10^{22}$ \cden) with the filtered and unfiltered data agreeing within $\lesssim 2$ K toward these regions.

Temperatures along the edges have $T \gtrsim 25$ K.  At such warm temperatures, our SEDs are not fit reliably with $\lambda \ge 160$ \um.  Additional, short-wavelength data at $\lesssim 100$ \um\ are needed to constrain the fits better.  Such an SED would also require more complicated models than our simplified modified blackbody functions to account for additional processes such as protostellar heating or non-equilibrium emission from very small dust grains \citep[e.g.,][]{Schnee08}.  Since such modeling is beyond the scope of this analysis, we instead excluded these warm regions.

In general, the dust temperatures appear to be most similar for the densest material from the unfiltered and filtered datasets.  In contrast, their respective column densities show more significant variations, with lower values obtained from the filtered data than from the unfiltered data. For example, column densities of $> 2 \times 10^{22}$ \cden\ cover an area of $\sim 30$ arcmin$^2$ with the unfiltered data and only $\sim 5$ arcmin$^2$ with the filtered data.  Moreover, the peak column density decreases from $1.5 \times 10^{23}$ \cden\ to $1.0 \times 10^{23}$ \cden\ between the unfiltered and filtered results, respectively.  These differences mean that we must be cautious when comparing column densities between studies with varying levels of spatial recovery.

\citet{Salji15} used flux ratios of 450 \um\ and 850 \um\ data at $\sim 14$\arcsec\ resolution from SCUBA-2 to determine temperatures and column densities for all of Orion A (including OMC 2/3) with the assumption of $\beta = 2$.   Comparing their results to ours from the \emph{filtered} data, \citet{Salji15} obtained cooler temperatures by $\sim 5$ K and higher column densities by factors of several to an order of magnitude toward the densest material in the OMC 2/3 filament.  These differences may be due to the assumption of $\beta = 2$.  If we assume $\beta = 2$ with the filtered data, the resulting temperatures decrease by $\sim 7$ K and the column densities increase by several factors, both in better agreement with \citet{Salji15}.  Since we found $\beta \simeq 1.6$ with the filtered data, assuming $\beta = 2$ will have a significant effect on both temperature and column density (see Sect. \ref{largeScaleSection}).

For those regions with well-defined SEDs ($T < 25$ K), we found a median temperature of 22 K and a median column density of $2.4 \times 10^{22}$ \cden.  These values agree well with the corresponding median values of 22 K / $2.4 \times 10^{22}$ \cden\ and 21 K / $2.6 \times 10^{22}$ \cden\ from \citet{Lombardi14} and \citet{StutzKainulainen15}, respectively, even though both of these studies used \emph{Herschel} data alone and fixed dust properties. \citet{Lombardi14} adopted $\beta \simeq 1.8$ based on the \emph{Planck}-determined dust emissivities for Orion A, whereas \citet{StutzKainulainen15} adopted the dust opacities from \citet[][OH5]{Ossenkopf94} that represent the expected dust grain properties for molecular clouds; we note that this particular dust opacity law has $\beta \simeq 1.8$.  Since both studies used values of $\beta$ similar to our median value, it is unsurprising that our respective datasets match well.  In general, our temperatures and column densities agree within $\sim$ 15\%\ with those in both \citet{Lombardi14} and \citet{StutzKainulainen15}. Our analysis, however, was restricted to OMC 2/3, and as such, we cannot comment on the results elsewhere in Orion A.

\begin{figure}[h!]
\centering
\includegraphics[scale=0.55]{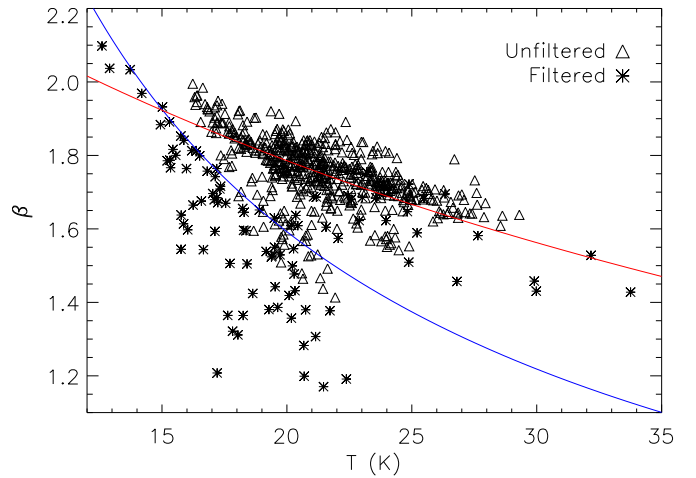}
\caption{Relationship between temperature and $\beta$ in OMC 2/3.  Triangles show the results from our unfiltered data and stars the results from the filtered data for pixels with $\NHH > 2 \times 10^{22}$ \cden.  The red curve shows the $T-\beta$ relation from \citet{Dupac03} and the blue curve shows the relation from \citet{Desert08}.}\label{beta_temp}
\end{figure}

Figure \ref{beta_temp} shows the temperature-$\beta$ relations as determined by our unfiltered and filtered \emph{Herschel}+GISMO data.  Both distributions show an anticorrelated relation as seen in other studies \citep[e.g.,][]{Dupac03, Desert08, Planck_beta2, Juvela15}, with a population of relatively warm ($\sim 20$ K) dust with low-$\beta$ values ($< 1.6$) that deviate from the main anticorrelation.  These pixels are associated exclusively with the MMS 6 core.

Figure \ref{beta_temp} includes $\beta$-temperature relationships from PRONAOS \citep{Dupac03} and Archeops \citep{Desert08} for comparison.  Excluding the MMS 6 data, our unfiltered data agree better with the anticorrelation from \citet{Dupac03}, even though their data were restricted to $\lambda < 600$ \um.  In contrast, the correlation from \citet{Desert08} included 2 mm data, as in this study, but did not fit our observations.  \citet{Desert08}, however, measured $\beta$ and temperature for individual clumps, whereas in the present study and in \citet{Dupac03}, $\beta$ and temperature were measured across an entire map, pixel by pixel.  Thus, the $\beta$-temperature relation may differ between compact sources and the larger-scale diffuse cloud, which might reflect a temperature dependence in $\beta$ itself or changes in these properties that are due to dust grain evolution \citep[e.g.,][]{Mennella98, Boudet05}.   

The anticorrelation in Fig. \ref{beta_temp}, however, might reflect the degeneracies between temperature and $\beta$ in the presence of noise \citep[e.g.,][]{Shetty09_noise, Juvela13} or from temperature variations along the line of sight \citep[e.g.,][]{Shetty09, Ysard12}.  In particular, $\chi^2$ analyses (such as those presented in this study) have been shown to produce skewed $\beta$-temperature relationships whereas more thorough Bayesian analyses have reproduced models more reliably  \citep{Kelly12, Juvela13, Juvela15}.  Nevertheless, the inclusion of long wavelength data can improve the reliability of SED fits via $\chi^2$ methods \citep{Juvela13}, and indeed, our \emph{Herschel}+GISMO data span a wide range in wavelengths, covering both the SED peak and the Rayleigh-Jeans tail.  

We tested the ability of the \emph{Herschel}+GISMO data to produce reliable measures of $\beta$ and temperature using model SEDs.  We generated mock fluxes corresponding to emission with $T=20$ K and $\beta = 1.7$, typical of what is seen in the OMC 2/3 region.  Then, we fitted the mock fluxes with modified blackbody functions assuming 10 K $\le T \le$ 80 K and 0 $\le \beta \le$ 5.0, assuming 10 \%\ calibration uncertainties in each band and the observed noise levels.

\begin{figure}[h!]
\includegraphics[scale=0.54]{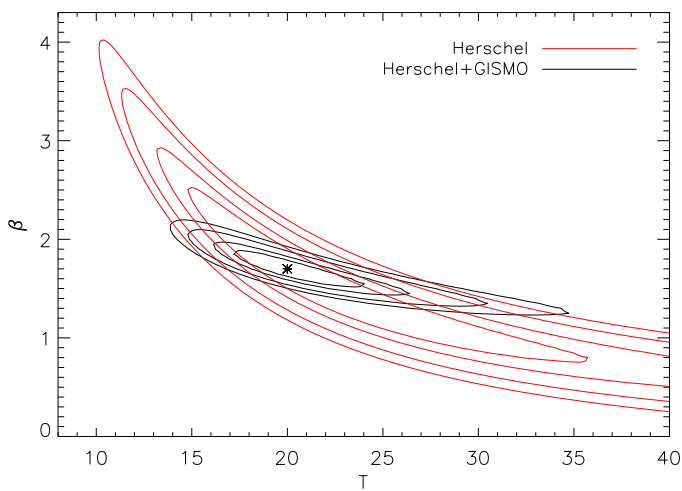}
\caption{Reduced $\chi^2$ contours from SED-fitting to \emph{Herschel}+2mm data (black) and \emph{Herschel}-only data (red) for mock data corresponding to $T = 20$ K and $\beta = 1.7$.  Contours correspond to reduced $\chi^2$ values of $\chi^2_{red} = 0.5,$ 1.0, 2.0, and 3.0, assuming four degrees of freedom for the \emph{Herschel}+2mm data and three degrees for the \emph{Herschel}-only data.  The star shows the input temperature and $\beta$ values from our model.  }\label{contour_plot}
\end{figure}

Figure \ref{contour_plot} shows distributions of reduced $\chi^2$ produced by each temperature-$\beta$ pair based on SED fits using either the \emph{Herschel}+GISMO bands and or the \emph{Herschel} bands alone.  Both distributions show curvatures that are indicative of the temperature-$\beta$ degeneracy from the calibration and noise uncertainties, although the \emph{Herschel}+GISMO case does show much tighter constraints for both parameters.  Additionally, the \emph{Herschel}+GISMO case recovered the input $\beta$ and temperature values from our model, whereas the \emph{Herschel}-only fits were skewed to warmer temperatures and lower $\beta$ indices.  Even with larger uncertainties at 2 mm that represent the errors in our large-scale 2 mm offsets (see Sect. \ref{largeScaleSection}), we recovered the input model $\beta$ and temperature values with the \emph{Herschel}+GISMO bands. Thus, the SEDs are well constrained by our \emph{Herschel} and 2 mm data.

\subsection{Implications for future analyses}\label{futureSection}

Most previous studies that combined \emph{Herschel} data with longer wavelength ground-based observations have been restricted to compact sources \citep[e.g.,][]{Maury11, Pezzuto12, Stutz13} or to filtering out the large-scale emission from the \emph{Herschel} data \citep[e.g.,][]{Sadavoy13, Wang15, Chen16}.  Conversely, \citet{Stutz10} combined \emph{Herschel} and ground-based data for a Bok globule without applying any filtering or offsets.  Adding a fixed offset (at 20\%\ of the peak flux) to ground-based data did not affect their results, suggesting that  the missing-large scale emission may not be significant for isolated Bok globules.  Other studies have used \emph{Planck} observations \citep{Csengeri16} or fixed offsets \citep{Forbrich15} to recover the large-scale emission from ground-based data and characterize the dust properties in clouds.

Using the technique outlined in Sect. \ref{largeScaleSection}, we can combine ground-based data with \emph{Herschel} data without filtering or assuming a fixed offset.    Our technique, however, is best used with ground-based observations at wavelengths far on the Rayleigh-Jeans tail of the SED.  For example, earlier tests with SCUBA-2 data at 850 \um\ were unable to constrain the SED fits clearly using a grid of offsets at 850 \um\ \citep{Sadavoy13}.  Since our GISMO 2 mm data are farther along the Rayleigh-Jeans tail, we have better leverage to constrain $\beta$ and the large-scale structure.

\section{Conclusions}\label{conc}

We have measured the dust emissivity index, $\beta$, across the high-mass star-forming OMC 2/3 filament using \emph{Herschel} $160 - 500$ \um\ and GISMO 2 mm observations and several methods to combine these datasets.  In particular, we produced the most complete and robust $\beta$ map of OMC 2/3 at 36\arcsec\ ($\sim$ 15000 AU) resolution.  Our main conclusions are listed below.

\begin{enumerate}

\item Using a technique that was first proposed in \citet{Sadavoy13}, we determined $\beta$ and recovered the missing large-scale emission from the 2 mm data simultaneously.  We tested our technique against filtering the large-scale emission from the \emph{Herschel} data and examining individual compact sources themselves.  We found that our  technique is consistent with these additional analyses and also with observations from \emph{Planck}.
\\[0pt]

\item We found that $\beta > 1.2$ everywhere in OMC 2/3, with most of the filament having $\beta = 1.7-1.8$.   The lowest $\beta$ indices were consistently associated with the MMS 6 core, which may indicate unique dust grain properties toward this object.  Nevertheless, the SEDs associated with MMS 6 are complicated by it being a protobinary system, and the lower $\beta$ values might reflect imperfect SED fits based on single-temperature modified blackbody functions.
\\[0pt]

\item We did not reproduce the very low indices of $\beta \simeq 0.9$ found by \citet{Schnee14}.  Instead, we found that their 3.3 mm observations appear to be elevated by roughly a factor of 2, which biased their measured $\beta$ indices to such low values.  These elevated fluxes may indicate significant contamination in this band or deviations from a single power law in the opacity curve.  At this time, we cannot determine the cause.  \\[0pt]

\item The 2 mm observations provided a strong constraint for the SED fits, resulting in relatively reliable measurements of $\beta$ and temperature over the \emph{Herschel} data alone.  Like many other studies, we found an anticorrelation between temperature and $\beta$.    
\end{enumerate}

We determined $\beta$ across the OMC 2/3 filament without filtering the \emph{Herschel} data or assuming a diffuse background offset value.  Our technique is reliable and easily applicable to other clouds observed with \emph{Herschel} and similarly long-wavelength, ground-based facilities.  Future analyses of dust properties in molecular clouds can use this method in lieu of removing the large-scale emission that made \emph{Herschel} submillimeter data so unique.

\vspace{1cm}
\begin{acknowledgements}
We thank the anonymous referee for providing helpful comments and suggestions.   This work was possible with funding from the Natural Sciences and Engineering Research Council of Canada PDF award.      The authors thank I. Hermelo, A. Kov{\'a}cs,  and J. Staguhn for valuable assistance with the GISMO observations and data reduction, and T. Stanke for the MAMBO observations of OMC 2/3.   IRAM is supported by INSU/CNRS (France), MPG (Germany) and IGN (Spain).   This project used data from the \emph{Herschel} Science Archive and the \emph{Planck} Legacy Archive.
\end{acknowledgements}

\bibliographystyle{aa}
\bibliography{references}


\begin{appendix}

\section{Filtering the Herschel data} \label{appendixFT}

Since GISMO is a ground-based instrument, its data are filtered on scales similar to the array footprint to remove the effects of a variable atmosphere.  The \emph{Herschel} data, however, are space-based and are not subject to the same level of filtering.  In Sect. \ref{filterSection}, we fitted SEDs to the \emph{Herschel}+GISMO data using similarly filtered \emph{Herschel} maps.  We describe this filtering process in more detail here.

\begin{figure}[h!]
\includegraphics[scale=0.35]{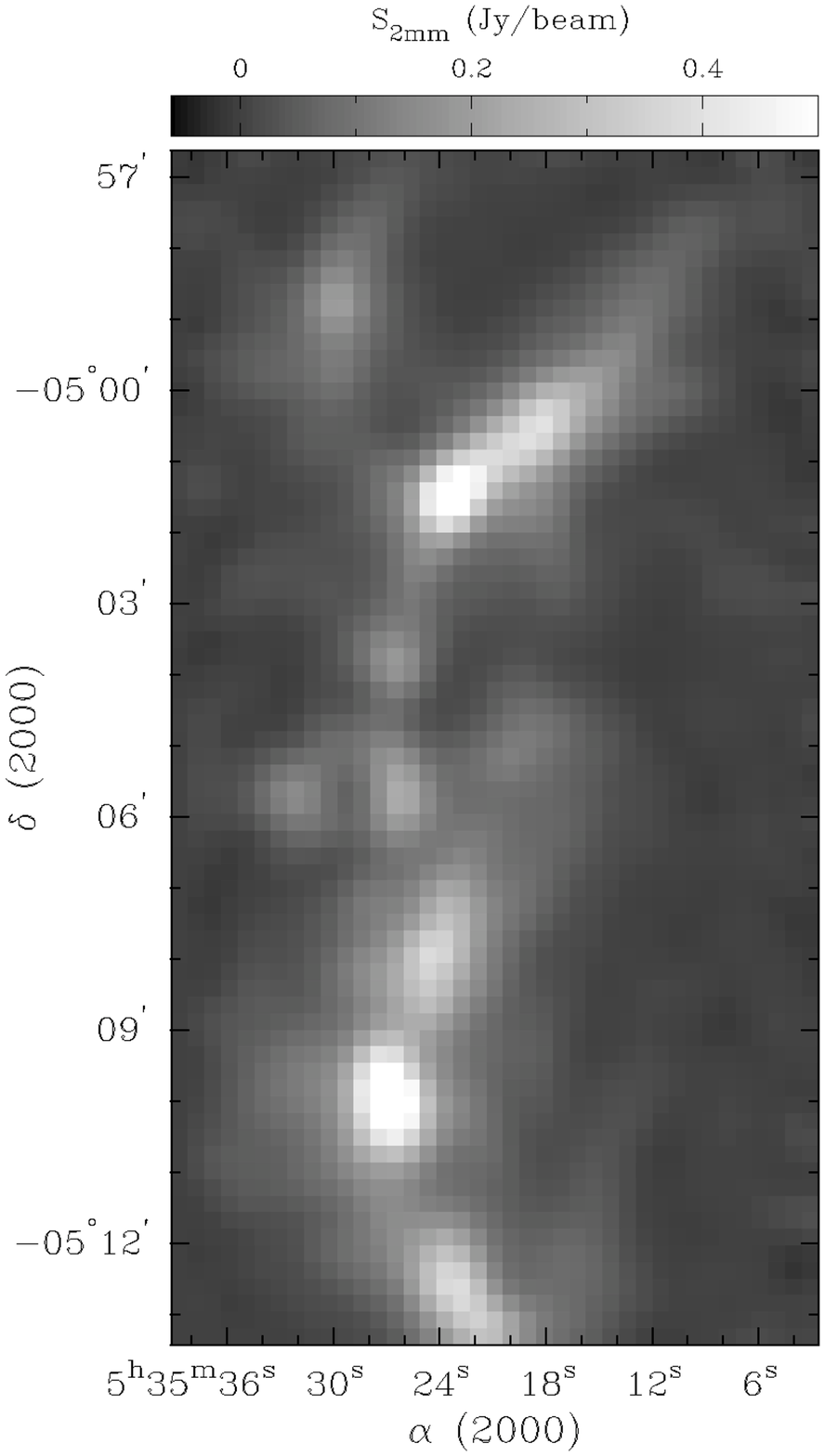}
\includegraphics[scale=0.655,trim=0pt -0.9cm 0cm 5mm,clip=true]{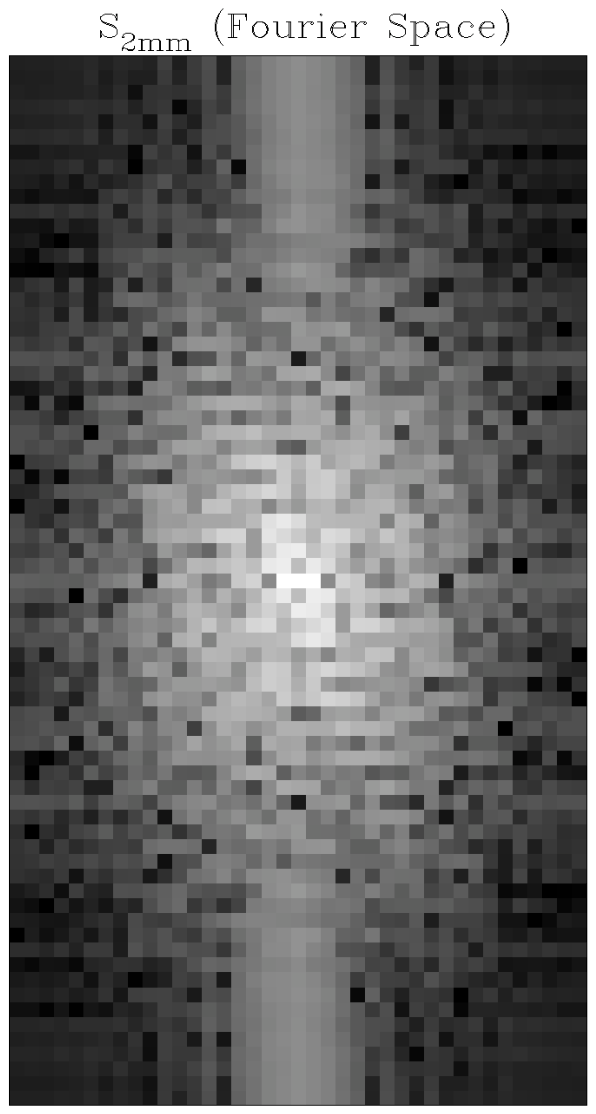}\\[2pt]
\includegraphics[scale=0.54]{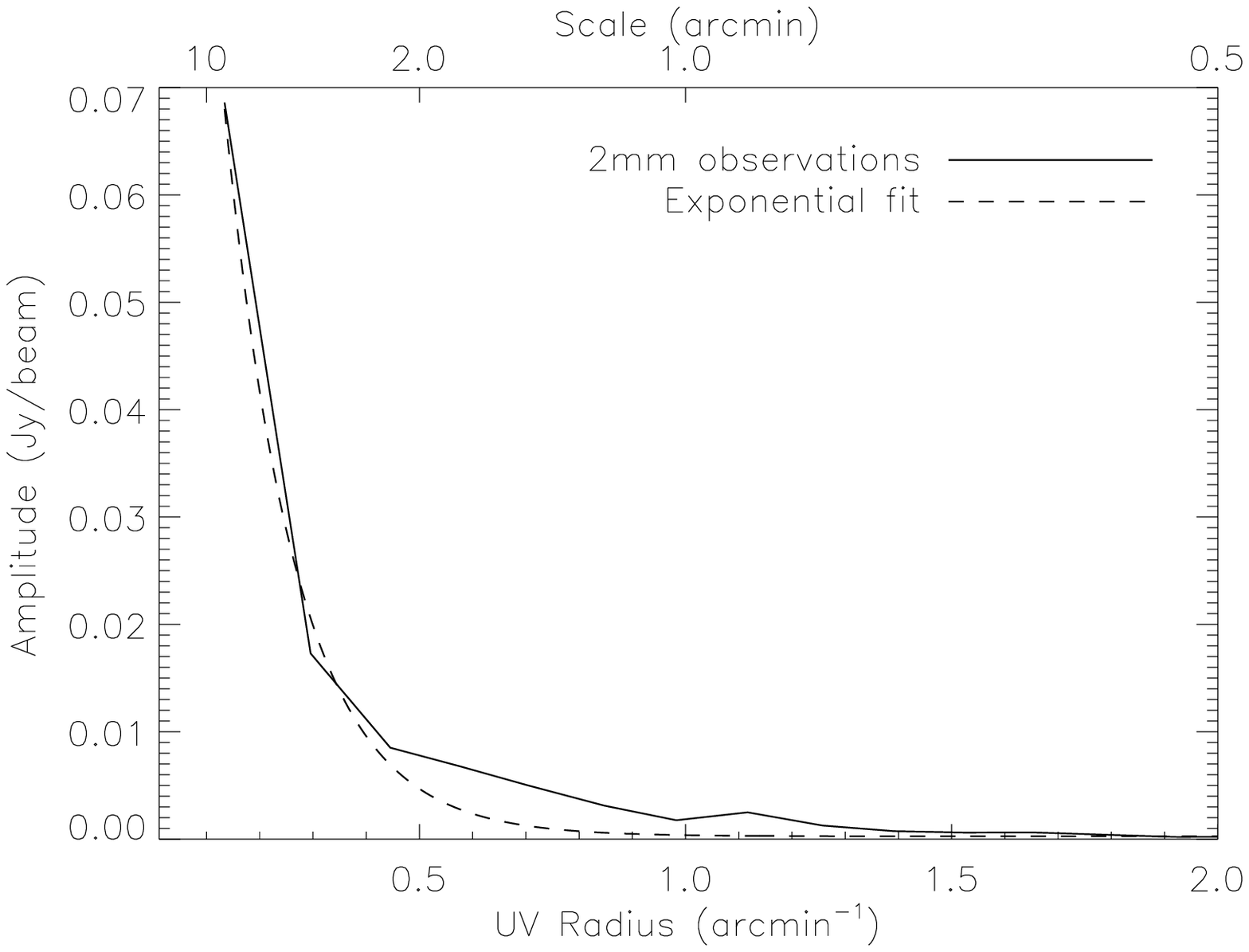}
\caption{\emph{Top left:} GISMO 2 mm observations of OMC 2/3 at 36.3\arcsec\ resolution.  \emph{Top right:} Fourier transform of the convolved GISMO 2 mm data. \emph{Bottom:} Radial amplitude profile of the GISMO 2 mm data.  The dashed line shows the best-fit exponential function to the data.}\label{fft_2mm}
\end{figure}

Following \cite{Wang15}, we removed the large-scale emission from the \emph{Herschel} data by suppressing low spatial frequencies similar to a highpass filter.  We determined which scales to suppress from the GISMO 2 mm data themselves to ensure that we filtered out similar spatial scales.  In Fig. \ref{fft_2mm}, we show the GISMO 2 mm data at 36\arcsec\ resolution and the corresponding Fourier transformation.  The bottom panel of Fig. \ref{fft_2mm} shows the radial amplitude profile of the GISMO 2 mm frequency domain data.  The radial profile was constructed from azimuthal averages in increasing bin sizes of one pixel.  We fitted this amplitude profile with an exponential function, $f_{exp}(r)$ (dashed curve).  This exponential function represents the relative sensitivity to each scale.  Therefore we used $1 - f_{exp}^{norm}(r)$ as our mask, where $f_{exp}^{norm}(r)$ is normalized.  Thus, at short uv radii, our mask reaches zero (e.g., filters out the large-scale emission), whereas our mask reaches one at large uv radii.

\begin{figure}[h!]
\includegraphics[scale=0.65]{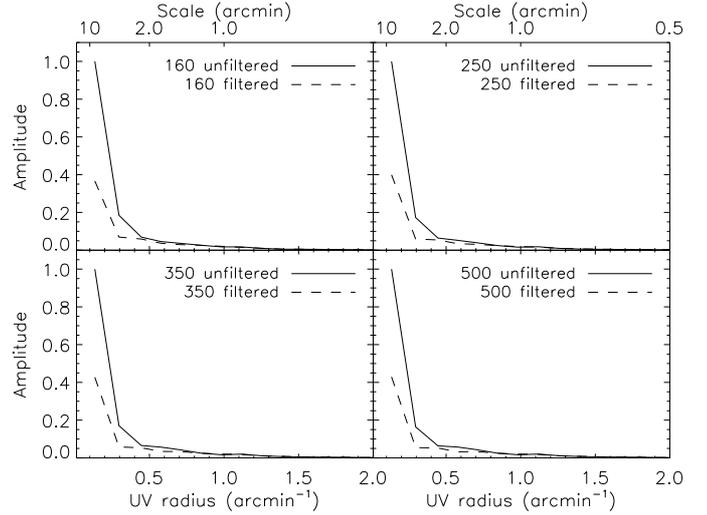}
\caption{Radial amplitude profiles for the \emph{Herschel} bands for a $32 \times 64$ pixel region of OMC 2/3.  The solid curves show the observed (unfiltered) amplitude profiles, normalized by the peak value.  The dashed curves show the amplitude profiles of our filtered data following our large-scale masking.  The filtered profiles were scaled by the same normalization factors as the unfiltered profiles.  All data are relative to the 36\arcsec\ convolved \emph{Herschel} data.}\label{amp_herschel}
\end{figure}

We multiplied our exponential mask with the \emph{Herschel} data in frequency space, and then converted the products back to the image plane.  Figure \ref{amp_herschel} compares the unfiltered and filtered amplitude profiles for the \emph{Herschel} bands following our technique.  In general, we find that the filtered and unfiltered \emph{Herschel} amplitude profiles agree well for spatial scales $\lesssim$ 2$\arcmin$, in line with the expectation that emission is recovered at levels of $\sim 90$\%\ at $\sim 2$\arcmin.  At larger spatial scales (e.g., $> 5$\arcmin), the amplitude decreases considerably because of filtering.

\begin{figure}[h!]
\includegraphics[scale=0.65]{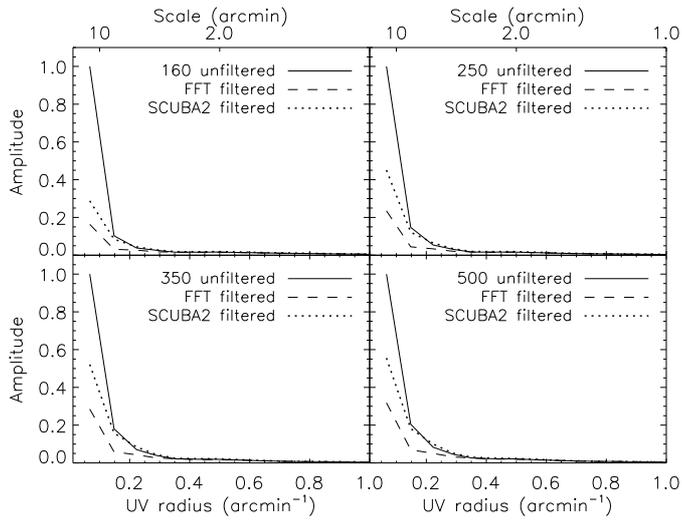}
\caption{Same as Fig. \ref{amp_herschel}, but for a $64 \times 64$ pixel region of B1 in Perseus \citep{Sadavoy13}.  The dotted curves show the amplitude profiles for the corresponding \emph{Herschel} maps that were filtered by the SCUBA-2 pipeline.}\label{amp_herschel_b1}
\end{figure}

As an additional test of our method, we applied this Fourier-space filtering technique to the \emph{Herschel} data of Perseus B1.  These \emph{Herschel} maps were filtered also through the SCUBA-2 pipeline, which is expected to reflect  the SCUBA-2 filtering process more reliably \citep[see][]{Sadavoy13}.  Figure \ref{amp_herschel_b1} compares the \emph{Herschel} amplitude profiles for the unfiltered observations, the Fourier-filtered data, and the SCUBA-2 pipeline filtered data.  For the Fourier-filtered data, we used the original SCUBA-2 850 \um\ amplitude profile for B1 in a similar manner as described above for the GISMO data.  The Fourier-filtering method generally removed more emission  than the SCUBA-2 pipeline on scales $\gtrsim 5\arcmin$.  Nevertheless, both methods agree well for scales of $\lesssim 3$\arcmin, where we expect most emission to be recovered.  Thus, our Fourier-filtering technique appears to give a reasonable approximation of the filtering from ground-based facilities.

We note that a similar exercise to filter out emission corresponding to the MUSTANG pipeline \citep[e.g., for comparison to][]{Schnee14} is inadvisable.   The MUSTANG pipeline recovered 3.3 mm emission only to $\sim 1\arcmin$. Since the SPIRE resolutions, particularly at 500 \um, are comparable to this largest recoverable scale, filtering the \emph{Herschel} data with either the MUSTANG pipeline \citep[e.g., as in][]{Schnee14} or with our Fourier transform method is highly suspect.  In particular, the inherent resolution differences in the SPIRE data may result in a disproportionate fraction of emission being filtered.

\end{appendix}

\end{document}